\documentclass[twocolumn,prl,aps,showpacs]{revtex4}
\usepackage{graphics}% Include figure files
\usepackage{dcolumn}% Align table columns on decimal point
\usepackage{bm}% bold math
\usepackage{epsfig}
\usepackage{pictex}

\begin{document}

\title{Non-Fermi liquid behavior in nearly charge ordered layered metals}
\author{J. Merino$^1$, A. Greco$^2$, N. Drichko$^{3,4}$ and M. Dressel$^3$}
\affiliation{$^1$Departamento de F\'isica Te\'orica de la Materia
Condensada, Universidad Aut\'onoma de Madrid, Madrid 28049, Spain \\
$^2$ Facultad de Ciencias Exactas Ingenier\'ia y Agrimensura e Instituto de F\'isica Rosario, Rosario, Argentina \\
$^3$ 1.\ Physikalisches Institut, Universit\"at Stuttgart, D-70550
Stuttgart, Germany\\
$^4$Ioffe Physico-Technical Institute, 194021 St. Petersburg, Russia}
\date{\today}
\begin{abstract}
Non-Fermi liquid behavior is shown to occur in two-dimensional
metals which are close to a charge ordering transition driven by
the Coulomb repulsion. A linear temperature dependence of the
scattering rate together with an increase of the electron effective mass
occur above $T^*$, a temperature scale much smaller than the Fermi
temperature. It is shown that the anomalous temperature dependence of the
optical conductivity of the quasi-two-dimensional organic metal
$\alpha$-(BEDT-TTF)$_2$$M$Hg(SCN)$_4$, with $M$=NH$_4$ and Rb, above
$T^*=50-100$ K, agrees qualitatively with predictions for the
electronic properties of nearly charge ordered two-dimensional
metals.
\end{abstract}
\maketitle

Charge ordering phenomena appear in various strongly correlated
systems such as magnetite (Fe$_3$O$_4$) \cite{Walz}, rare earth
manganites \cite{Chen}, the quasi-two-dimensional organic
conductors $\theta$-, and $\alpha$-(BEDT-TTF)$_2$$X$
\cite{Seo,Dressel04} and Na$_x$CO$_2$ \cite{Ong}. The relevance of
charge ordering (CO) to the superconductivity appearing in hydrated
samples\cite{Takada} of Na$_{0.35}$CO$_2$ \cite{Lee} and the
$\theta$-, $\beta''$-(BEDT-TTF)$_2$$X$ layered organic compounds\cite{Merino2}
has also been recently pointed out. Unconventional features observed in
the optical conductivity \cite{Dressel} of the quasi-two-dimensional organic compounds:
$\alpha$-(BEDT-TTF)$_2$$M$Hg(SCN)$_4$ and electron
Raman scattering\cite{Lemmens} in Na$_x$CO$_2$ have been
interpreted in terms of a metal close to CO. This
is similar to the situation in high-T$_c$ superconductors in which
the proximity of antiferromagnetism and superconductivity has led
to an intense activity in understanding the properties of metals
close to an antiferromagnetic instability\cite{Moriya}. Anomalous
metallic properties such as the opening of a pseudogap
\cite{Kampf} and a linear temperature dependence of the
resistivity \cite{Hlubina} appear as a consequence of the
development of strong short-range antiferromagnetic correlations.

In this Letter we analyze the electronic properties of metals
close to a charge-order instability driven by the off-site Coulomb
repulsion. We show that contributions to the single-particle
self-energy due to the interaction of electrons with charge
fluctuations {\it increase} as the temperature increases. This
effect leads to effective masses increasing with temperature;
opposite to the behavior expected in nearly magnetically ordered
metals or in metals with strong electron-phonon interaction. This
anomalous $T$-dependence of the effective mass is due to melting
of CO with {\it decreasing} temperature appearing in Wigner-type
transitions {\it i.e.} CO transitions driven by the long range
part of the Coulomb repulsion. Remarkably, we find that the
temperature dependence of the optical conductivity measured on
$\alpha$-(BEDT-TTF)$_2$$M$Hg(SCN)$_4$ with $M$=NH$_4$ and Rb above
50 K, fits qualitatively that of a two-dimensional metal close to
charge-ordering.

We consider the simplest model that makes CO possible
due to competition between kinetic and Coulomb energies, {\it
i.e.} the extended Hubbard model:
\begin{eqnarray}
H &=& \sum_{<ij>,\sigma}\;(t_{ij}\; c^\dag_{i\sigma} c_{j \sigma} +
h.c.) + U \sum_i n_{i\uparrow} n_{i \downarrow}
\nonumber  \\
&+&\sum_{<ij>} V_{ij} n_i n_j -\mu \sum_{i \sigma} n_{i \sigma},
\label{hamilt}
\end{eqnarray}
which describes fermions in a lattice with an on-site Coulomb
repulsion $U$, a nearest-neighbors Coulomb repulsion $V_{ij}$ and
a hopping matrix element $t_{ij}$ which expresses the hopping
processes between nearest-neighbors sites of the lattice. The
$c^\dag_{i\sigma} (c_{i \sigma}) $ denote creation (annihilation)
operators for the electron with spin $\sigma$ at the $i$-th site,
respectively, and $n_i=n_{i\uparrow}+n_{i\downarrow}$ where
$n_{i\sigma}=c^\dag_{i \sigma} c_{i \sigma}$. We assume the
simplest two-dimensional case of the square lattice {\it i.e.}
$t_{ij}=-t$ and $V_{ij}=V$ with half a hole per site (quarter
filling), $n=<n_i>=3/2$.  We choose $t<0$ which gives a hole-like
Fermi surface more appropriate for the description of $\alpha$ and
$\theta$-type layered molecular conductors.

The ground state phase diagram of model (\ref{hamilt}) at
$1/4$-filling has been studied by means of Hartree-Fock
\cite{Seo}, exact diagonalization \cite{Calandra}, large-$N$
Hubbard operator theory\cite{Merino1} and slave bosons
\cite{McKenzie}. A robust feature of the CO transition driven by
$V$ is its reentrant behavior with decreasing temperature (from a
metal to a charge-ordered state back to a metal) which has been
predicted by dynamical mean-field theory \cite{Bulla}, slave boson
theory\cite{Merino2} and finite-$T$ Lanczos
diagonalization\cite{Hellberg}. The random phase
approximation (RPA) on the extended Hubbard model\cite{Ogata}
agrees with these approaches \cite{footnote1} recovering, in particular, the
reentrant character of the CO transition. Little is known about 
the electronic properties of metals close to CO so we cover this 
gap by calculating the $T$-dependence of one-electron properties 
based on RPA.

\begin{figure}
\begin{center}
\epsfig{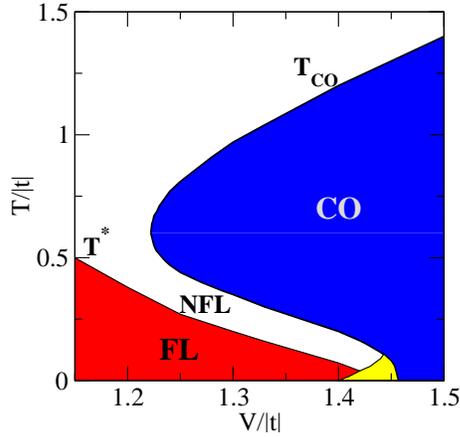}
\end{center}
\caption{(Color online) Non-Fermi liquid behavior in a metal
close to charge ordering. The $T$-$V$ phase diagram of the
$3/4$-filled extended Hubbard model for $U=2|t|$ is
shown. At $T=T_{\text{CO}}$ a transition to the checkerboard charge
ordered (CO) phase occurs (blue region). The low temperature
scale, $T^*$, separates Fermi liquid (FL) behavior at low
temperatures from non-Fermi liquid (NFL) above $T^*$.
A transition to a 2${\bf k_F}$-CDW occurs close to CO
(yellow region).}
\label{fig1}
\end{figure}

The $T$-$V$ phase diagram for $U=2|t|$ plotted in Fig.~\ref{fig1}
summarizes our main results. At $T=0$ a transition to a 2k$_F$-CDW
takes place at about $V_c \approx 1.41 |t|$. The temperature scale
$T_{\text{CO}}$ is the critical temperature for the checkerboard
CO\cite{Ogata} transition.  Non-Fermi liquid behavior occurs above
$T^*$ which evolves into conventional metallic behavior for
$T<T^*$. Decreasing temperature close to the transition drives the
system from a uniform metal to a CO state which transforms back to
a metal as temperature is further increased. This reentrant
behavior of the CO transition is responsible for the anomalous
effective mass increasing with temperature above $T^*$ as
discussed below.

\begin{figure}
\begin{center}
\epsfig{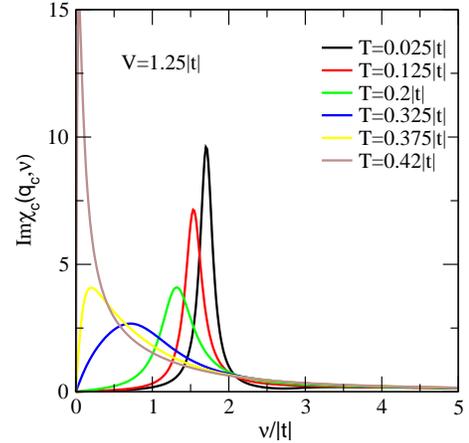}
\end{center}
\caption{(Color online) Softening of the ${\bf q_c}=(\pi,\pi)$ mode induced by
temperature close to the charge-ordering transition for $U=2|t|$
and $V=1.25|t|$. It is this $T$-dependence of the
charge susceptibility which amplifies the self-energy corrections
with increasing temperature leading to anomalous thermal increase of
effective masses.}
\label{fig2}
\end{figure}

The reentrant behavior of the CO transition appearing
in Fig. \ref{fig1} can be understood from the precise form of
the RPA charge susceptibility in model (\ref{hamilt}), which reads
\begin{equation}
\chi_c( {\bf q},i \nu_n )={\chi_0({\bf q}, i \nu_n) \over 1+V({\bf
q})\chi_0( {\bf q}, i \nu_n) }
\label{chic}
\end{equation}
where $V({\bf q})=U/2+2V(\cos q_x +\cos q_y)$ (lattice spacing $a = 1$) 
and $\chi_0({\bf q}, i \nu_n)$ is the non-interacting positive charge susceptibility
which includes the two spin species and $\nu_n$ are bosonic
Matsubara frequencies. Fixing $U=2|t|$, the static charge
susceptibility $\chi_c({\bf q},0)$ diverges for ${\bf q} \approx
2{\bf k_F}=(2.4,2.4)$ at $V_c$. Increasing temperature shifts the
instability to ${\bf q_c}=(\pi,\pi)$ because then $V({\bf q})$
dominates in the denominator of Eq. \ref{chic} driving the system
into the checkerboard charge ordered state\cite{Ogata}. This leads
to a softening of the ${\bf q}_c=(\pi,\pi)$ mode with temperature
as shown in Fig.~\ref{fig2} where the imaginary part of the charge
correlation function is plotted in the reentrant region of the
transition ($V =1.25|t|$). It is this softening induced by
temperature which finally amplifies the contributions to the
one-electron self-energy.

\begin{figure}
\begin{center}
\epsfig{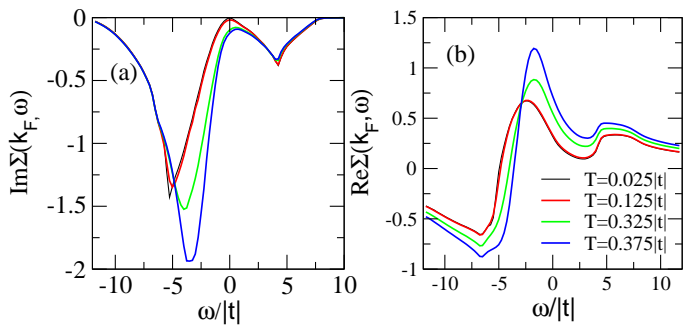}
\end{center}
\caption{(Color online) Temperature dependence of self-energy for $U=2|t|$ and
$V=1.25|t|$. In (a) the imaginary part of the self-energy is
plotted displaying the enhancement for $\omega <0$ with temperature,
while in (b) the real part shows the gradual
increase of the slope at the Fermi energy above $T^*$.} \label{fig3}
\end{figure}

Self-energy corrections induced by temperature, at the RPA level, read:
\begin{eqnarray}
\Sigma({\bf k}, i \omega_n) = {1 \over \Omega} \sum_{\bf q} V({\bf
q})^2  \Big\lbrace \int_0^{\infty} {{\rm d} \nu \over \pi}
\text{Im} \chi_c({\bf q}, \nu)
\nonumber \\
\Big \lbrack {n_B(\omega')+1-n_F(\epsilon_{\bf k-q}) \over i
\omega_n - \nu - \epsilon_{\bf k-q}} +
{n_B(\omega')+n_F(\epsilon_{\bf k-q}) \over i \omega_n + \nu -
\epsilon_{\bf k-q}} \Big\rbrack \Big\rbrace, \label{self}
\end{eqnarray}
with $\Omega$ being the volume. The imaginary and real parts of the
self-energy, evaluated at the Fermi momentum
${\bf k}_F=(1.2,1.2)$ with $U=2|t|$ and $V=1.25|t|$, are plotted in
Fig.~\ref{fig3}. As the temperature is increased, the imaginary
part of the self-energy increases for hole propagation ($\omega<0$)
due to the coupling to the $(\pi,\pi)$ charge fluctuating mode. 

The anomalous $T$-dependence of the self-energy close to CO
can be further explored by approximating the charge susceptibility with
\begin{equation}
\chi_c({\bf q \approx q}_c, i \omega) \approx {\chi_c({\bf q}_c)
\over \omega_{\bf q} - i\omega}, \label{chicap}
\end{equation}
with $\omega_{\bf q}=\omega_0(T) + B({\bf q} - {\bf q}_c)^2$ and
$\chi_c({\bf q}_c) = \chi_c({\bf q}_c,0)$. At $T=0$, $\omega_0(T) \rightarrow 0$
for the ${\bf q_c}=2{\bf k_F}$-type CDW as $V \rightarrow V_c$. At
finite-$T$, $\omega_0(T) \rightarrow 0$ as $T \rightarrow T_{\text{CO}}$
signalling the CO transition at finite temperatures.

\begin{figure}
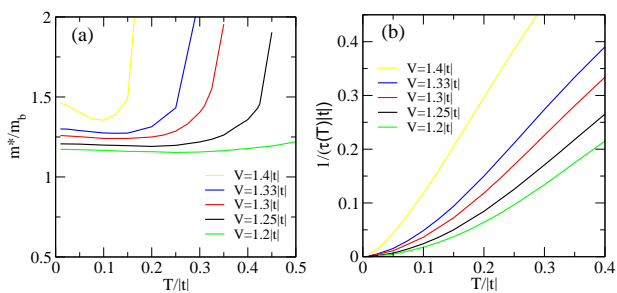

\begin{center}
\epsfig{file=fig4a.eps,width=4.cm,angle=0,clip=}
\epsfig{file=fig4b.eps,width=4.cm,angle=0,clip=}
\end{center}
\caption{(Color online) Anomalous electron effective mass
enhancement induced by temperature and non-Fermi liquid behavior
in a nearly charge ordered metal above $T^*$. In (a) the
temperature dependence of the effective mass enhancement is shown
for $U=2|t|$ and different $V$'s close to CO while in (b)
the $T$-dependence of the scattering rate shows a linear
$T$-dependence above $T^*$.} \label{fig4}
\end{figure}

The imaginary part of the self-energy at the Fermi energy, using
the approximate expression (\ref{chicap}) is
\begin{equation}
\text{Im}\Sigma({\bf k_F},0) \approx \chi_c({\bf q_c}) T \int_{FS}
{ {\rm d} {\bf q} \over (2 \pi)^2 |v_{{\bf k_F-q}}|}  {V({\bf
q})^2 \over \omega_{\bf q} } {\rm arctan}\left({T \over
\omega_{\bf q}}\right), \label{imsigma}
\end{equation}
with the integration taken over momenta on the Fermi surface. Within
RPA, the above expression is exact at low and high temperatures. The
scattering rate exhibits Fermi-liquid behavior: $1/\tau(T)\approx
A_2 T^2$, at low temperatures $T<<T^*$, while at large
temperatures, $T>>T^*$: $1/\tau(T) \approx A_1 T$ with the prefactors
\begin{equation}
A_i=\chi_c({\bf q_c}) \int_{FS} {d {\bf q} \over (2 \pi)^2 |v_{{\bf k_F-q}}|} {V({\bf q})^2 \over
\omega_{\bf q}^i},
\label{pref}
\end{equation}
which are enhanced as $V \rightarrow V_c$. The ${\bf q_c}=2{\bf
k_F}$ mode is responsible for the rapid increase in the scattering
rate slope appearing in Fig. \ref{fig4}(b) as it connects
different points at the Fermi surface giving the dominant
contribution to the integration in Eq. (\ref{pref}).
Simultaneously $T^* \rightarrow 0$ as $V\rightarrow V_c$, so the
scattering rate behaves linearly down to very low temperatures
close to CO.

We finally analyze the effective mass enhancement using Eq.
(\ref{chicap}) in $m^*/m_b=1/Z=(1-\partial$Re$\Sigma({\bf
k_F},\omega)/\partial \omega) |_{w=0}$, where $m_b$ is the band mass.  At
$T=0$, the effective mass increases as $V \rightarrow V_c$ due to
the proximity to the $2 {\bf k_F}$-CDW instability. This
enhancement is $\sim \ln(1/\omega_{\bf q})$ with ${\bf q} \approx
2 {\bf k_F}$ and with temperature increase below $T^*$, a
suppression of the effective mass occurs which is $\sim
(T/\omega_{\bf q})^3\ln(T/\omega_{\bf q})$ characteristic also of
nearly magnetically ordered metals\cite{Moriya,Doniach}. This
behavior is more apparent in Fig. \ref{fig4}(a) when $V$ is
sufficiently close to $V_c$. However, this low temperature
decrease reverses at larger temperatures {\it i. e.} above $T^*$
electrons become {\it heavier} with temperature.
This results from the softening of the ${\bf q} \approx (\pi,\pi)$
modes as $T \rightarrow T_{\text{CO}}$ shown in Fig. \ref{fig2}.

\begin{figure}
\begin{center}
\epsfig{file=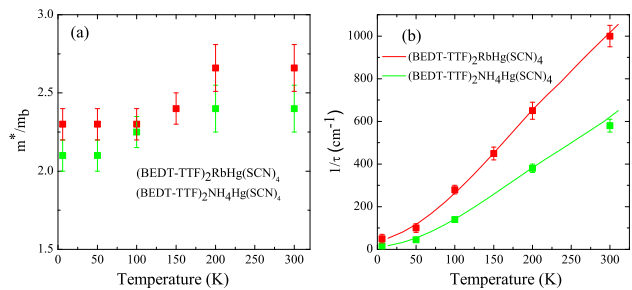,width=9.7cm,angle=0,clip=}
\end{center}
\caption{(Color online) Temperature dependence of 
one-electron properties obtained from optical
measurements on $\alpha$-(BEDT-TTF)$_2$$M$Hg(SCN)$_4$ for the
polarization parallel to the BEDT-TTF stacks. The thermal 
increase of effective mass enhancement (a) and the linear $T$-dependence of the
scattering rate (b) observed above $T>T^* \approx 50-100$~K
agrees qualitatively with RPA predictions of nearly two-dimensional 
charge ordered metals.}
\label{fig5}
\end{figure}

In order to compare calculations with experiments on
$\alpha$-(BEDT-TTF)$_2$$M$Hg(SCN)$_4$, let us first review some of
their properties. The $\alpha$-(BEDT-TTF)$_2$$M$Hg(SCN)$_4$
compounds with $M$=K, Tl and Rb are quarter-filled (with holes)
systems which display a density wave ground state\cite{McKenzie1}
below $T_{DW}=6-10$ K, whereas the $M$=NH$4$ compound is the only
member of the family which exhibits superconductivity at
$T_c=1$~K.  The low energy density wave state is attributed to
nesting of one-dimensional sections of the Fermi surface and is
rapidly degraded with temperature\cite{footnote2}. From our
reflectivity measurements on $\alpha$-(BEDT-TTF)$_2$$M$Hg(SCN)$_4$
compounds an analysis of the spectral weight and the width of the
zero-frequency contribution of the frequency dependent
conductivity yields the effective mass and scattering rate at
certain temperatures \cite{Dressel04,Drichko}. In Fig.~\ref{fig5}
the experimental findings are displayed. The scattering rates for
both salts show a linear temperature dependence in a broad
temperature range: $T>50-100$~K. Remarkably, around this
temperature a change in the $T$-dependence of the effective mass
can be also identified, showing an {\it increase} with
temperature\cite{footnote3}. This crossover temperature scale is about two orders
of magnitude smaller than the Fermi energy. From our theoretical
predictions of a metal close to a charge-ordering transition, the
linear $T$-dependence of the scattering rate and the increase in
the effective mass occur at a very small energy scale $T^*$ which
varies between $0.14|t|$ and $0.3|t|$ for $1.25<V/|t|<1.33$. For
typical values of the hopping matrix elements\cite{footnote4} for
$\alpha$-(BEDT-TTF)$_2$$M$Hg(SCN)$_4$, $t \approx 0.06$~eV,
theoretical estimates yield $T^*=0.14|t| \approx 80$~K for
$V=1.33|t|$ which is consistent with our
observations\cite{footnote5}. Experimentally it is also found that
the slope of the scattering rate of the NH$_4$ salt is smaller
than that of Rb. From a direct comparison with our theoretical
predictions (see Fig.~\ref{fig4}) we can conclude that the Rb
compound has a larger $V/|t|$ ratio than the NH$_4$ and is
effectively closer to the charge-ordering transition.  This
conclusion is consistent with the comparatively larger effective
mass enhancements observed for the Rb compound.

Other bosonic modes such as phonons can 
also lead to a linear temperature dependence in the scattering rate.
In this case the crossover temperature
scale set by the Debye temperature\cite{Kondo}, $\Theta \sim
200$~K, is larger than the experimental $T^*$. It is also  
expected that the coupling of electrons to phonons would lead 
to a {\it suppression} of the effective mass with temperature instead of
the {\it enhancement} experimentally observed.

In conclusion, we have shown that two-dimensional metals
sufficiently close to a charge-ordering transition display
non-Fermi liquid behavior including an {\it increase} in their
electronic effective mass with temperature. The effective mass
enhancement and the linear temperature dependence of the
scattering rate above $T^*$ predicted is in agreement with the optical
response experimentally observed in $\alpha$-(BEDT-TTF)$_2$$M$Hg(SCN)$_4$, 
indicating that they behave as nearly charge-ordered metals. Theoretical 
estimates of $T^*$ are comparable with the experimentally observed value of $T^* \sim
50-100$ K, which is much smaller than the Fermi temperature.
Additional probes such as angular resolved photoemission (ARPES)
should be used to obtain the full temperature dependence of the
electron self-energy. Strong coupling approaches should be used 
for determining the $T$-dependence of the self-energy and test 
the robustness of the present RPA analysis.

\acknowledgments J. M. acknowledges financial support from the
Ram\'on y Cajal program and MEyC under contract: CTQ2005-09385 in Spain. N.D.
thanks the Alexander von Humboldt-Foundation. The experimental work was
partially funded by the DFG. We acknowledge collaboration with J. Schlueter.


\begin{thebibliography}{}
\bibitem{footnote1} Critical temperatures, $T_c$, are typically overestimated by RPA.
However, RPA extensions such as the fluctuating exchange approximation
(FLEX) predict $T_c$'s in good agreement with experiments\cite{Moriya}.
\bibitem{footnote2} The Fermi surface of $\alpha$-(BEDT-TTF)$_2$MHg(SCN)$_4$
contains quasi-one-dimensional sections which, if included in
the calculation, would lead to similar anomalous $T$-dependence.
\bibitem{footnote3} 
The effective mass increase with $T$ above $\sim 50$ K is a robust
feature of our experimental data. However, limited accuracies
inherent to the analysis prevents us to extract its precise temperature
dependence.
\bibitem{footnote4} The hopping matrix element used here is appropriate for describing
the quasi-1D band that crosses the Fermi energy of
$\alpha$-(BEDT-TTF)$_2$MHg(SCN)$_4$ with bandwidth of 0.5 eV.
\bibitem{footnote5} As $V\sim U/2$, longer range Coulomb interactions
which can stabilize other Wigner-type charge ordered states different
from the checkerboard, may not be negligible.
\end{thebibliography}
\end{document}